# On-chip electro-optic tuning of a lithium niobate microresonator with integrated in-plane microelectrodes

**Min Wang,**[1,3] **Yingxin Xu,**[2] **Zhiwei Fang,**[1,3,4] **Yang Liao,**[1] **Peng Wang,**[1,3,4] **Wei Chu,**[1] **Lingling Qiao,**[1] **Jintian Lin,**[1,6] **Wei Fang**[2,7] **and Ya Cheng**[1,5,8]

[1] *State Key Laboratory of High Field Laser Physics, Shanghai Institute of Optics and Fine Mechanics, Chinese Academy of Sciences, Shanghai 201800, People's Republic of China*
[2] *State Key Laboratory of Modern Optical Instrumentation, College of Optical Science and Engineering, Zhejiang University, Hangzhou 310027, People's Republic of China*
[3] *University of Chinese Academy of Sciences, Beijing 100049, People's Republic of China*
[4] *School of Physical Science and Technology, ShanghaiTech University, Shanghai 200031, People's Republic of China*
[5] *State Key Laboratory of Precision Spectroscopy, East China Normal University, Shanghai 200062, China*
[6] *jintianlin@siom.ac.cn*
[7] *wfang08@zju.edu.cn*
[8] *ya.cheng@siom.ac.cn*

**Abstract:** We demonstrate electro-optic tuning of an on-chip lithium niobate microresonator with integrated in-plane microelectrodes. First two metallic microelectrodes on the substrate were formed via femtosecond laser process. Then a high-Q lithium niobate microresonator located between the microelectrodes was fabricated by femtosecond laser direct writing accompanied by focused ion beam milling. Due to the efficient structure designing, high electro-optical tuning coefficient of 3.41 pm/V was observed.

## 1. Introduction

High-Q crystalline microresonators have attracted tremendous attention for their broad range of applications in quantum electrodynamics, sensing, and optical signal processing [1]. Recently, several groups have demonstrated realization of free-standing on-chip microresonators in lithium niobate (LN) crystal of Q factors on the order of $10^6$ [2-5]. Among them, the maskless processing technique combining femtosecond laser direct writing with focused ion beam (FIB) milling allows rapid prototyping of LN microresonators of various sizes and geometries [2,3,6,7]. Besides, the femtosecond laser direct writing technique along can also provide functionality of straightforward integration of various functional microcomponents in a single substrate [8], which are highly in demand by either scientific research or industrial applications.

For most applications utilizing a microresonator, wavelength tuning is of vital importance [9]. The large electro-optic coefficient of LN has enabled fast and efficient tuning of hybrid silicon and LN ring microresonators with typical loaded Q factors on the order of $10^3 \sim 10^4$ [10, 11]. Electro-optic (EO) tuning of free-standing LN microresonators has recently been demonstrated by Bo et al. with external ITO electrodes on top of the microresonator [5]. Nevertheless, fully integrated wavelength tunable free-standing LN microresonators have not been demonstrated due to the difficulty in fabricating the tiny microelectrodes in the close vicinity to the microresonators. In this work, we show that using femtosecond laser direct writing the microelectrodes formed by electroless plating can be integrated with the microresonators in a straightforward fashion. We also demonstrate the EO tunability of the fabricated device reaching 3.41 pm/V.

## 2. Sample fabrication and experimental setup

In this work, commercially available ion-sliced X-cut $LiNbO_3$(LN) thin film with a thickness of 0.7 μm (NANOLN, Jinan Jingzheng Electronics Co., Ltd) was chosen for fabricating the on-chip EO tunable $LiNbO_3$ microdisk resonators. The LN thin film was bonded with a 500-μm thick LN substrate sandwiched by a $SiO_2$ layer with a thickness of 2 μm [12]. A Ti: sapphire femtosecond laser source (Coherent, Inc., center wavelength: 800 nm, pulse width: 40 fs, repetition rate: 1 kHz) was used for fabricating the microelectrodes and the LN microresonators. A variable neutral density filter was used to tune the average power of the laser beam. Objective lenses with different numerical apertures (NA 0.4 ~ 0.8) were used in the writing of the microelectrodes and LN microresonators. The femtosecond laser direct writing was performed by translating the LN thin film sample, which was mounted on a computer-controlled XYZ stage with 1-μm resolution, with respect to the stationary focal spot produced by the objective lens. A charged coupled device (CCD) connecting to the computer was installed above the objective lens to monitor the fabrication process in real time.

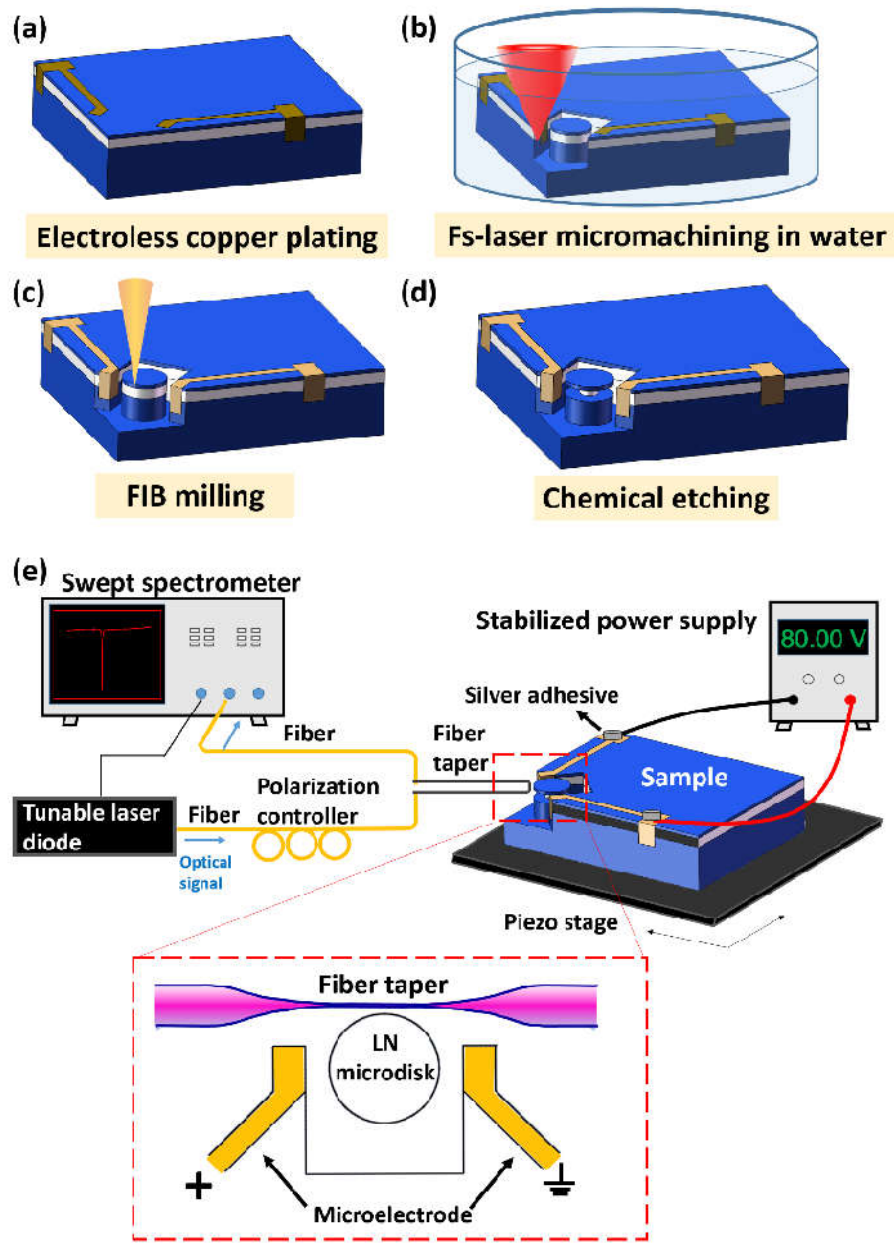

Fig. 1 (a-d) The processing flow of fabricating an on-chip electro-optic tunable LN microresonator with integrated in-plane microelectrodes: (a) Formation of two microelectrodes using femtosecond laser assisted selective electroless copper plating. (b) Water-assisted femtosecond laser ablation in LN substrate to form a cylindrical post. (c) Focused ion beam (FIB) milling to smooth the periphery of the cylindrical post. (d) Chemical wet etching of the fabricated sample to form the freestanding LN microdisk resonator. (e) Experimental setup for measuring the electro-optic tuning of the resonate wavelength in the $LiNbO_3$ microresonator.

The procedure of fabricating the integrated sample is schematically illustrated in Fig. 1(a)-(d). First, to realize the electro-optic tuning of the microresonator, two trenches with a depth of 6 μm were inscribed on the LN substrate surface using a 20X objective lens of NA 0.4 at an average power of 0.42 mW and a scan speed of 160 μm/s. Then, a pair of parallel embedded microelectrodes separated by 75 μm were produced by filling copper into the trenches using femtosecond assisted selective electroless copper plating. The details of the selective metallization can be found in Ref. 13. The sample was then completely cleaned with distilled water for removal of the electroless plating solution. The structure of the microelectrodes is schematically illustrated in Fig. 1(a).

Next, a cylindrical post was fabricated between the two planar electrodes by ablating the LN substrate immersed in water using femtosecond laser pulses focused with a 100X objective lens (NA 0.8), as shown in Fig. 1(b). The height of the cylindrical post is 13 μm. The periphery of the cylindrical post was then smoothed by focused ion beam (FIB) milling, as shown in Fig. 1(c). Finally, chemical wet etching, which selectively removes the silica underneath the LN thin film to form free-standing LN microdisk resonator, was conducted by immersing the fabricated structure in a solution of 2% hydrofluoric (HF) diluted with water, as shown in Fig. 1(d). The diameter of the LN microdisk was 26.34 μm. More details of the process flow of fabricating the LN microresonator can be found in Refs. 2 and 3.

The on-chip EO tuning of the LN microresonator was demonstrated using a setup schematically shown in Fig. 1(e). A narrow-band continuous-wave tunable diode laser (New Focus, Model 688-LN) was used to excite the whispering gallery mode (WGM) of the LN microresonator via a fiber taper fabricated by pulling a section of SMF-28 single mode fiber to a diameter of ~1 μm. By using an online fiber polarization controller, WGM modes in the microdisk with certain polarization were excited. A transient optical power detector (Lafayette, Model 4650) was used to measure the transmission spectra at the output end of the fiber taper. The measurement system could record the light signal over a 100 nm-wavelength-span with 0.5 pm wavelength resolution and 0.015 dB power accuracy in less than 1 second. As illustrated in the bottom inset of Fig. 1(e), the fiber taper was placed close to the microdisk boundary by adjusting the XYZ-piezo stage, coupling the laser light into and out of the LN microresonator. During the EO tuning, the fiber taper was placed in contact with the top surface of the LN microdisk to enhance the stability of the measurement. Selective excitation of a certain spatial mode can be achieved by adjusting the position of the fiber taper with respect to the microresonator. An open-loop piezo controller (Thorlabs, Model MDT693B) was used as the voltage supply for microelectrodes, which provided a variable voltage ranging from 0 V to 150 V with a resolution of 0.1 V.

## 3. Results and discussion

Figure 2(a) shows the top view of a fabricated device consisting of a high-Q LN microresontor and two microelectrodes. Figure 2(b) presents the scanning electron microscope (SEM) image of the freestanding microdisk. The sidewall of the microdisk is very smooth, however, a taper angle of approximately 10° can be observed which is caused by the FIB milling.

Characterization of the optical mode spectrum of the microresonator was accomplished with the fiber taper coupling method [14]. First, a coarse scan was performed in the range from 1560 nm to 1617 nm with a wavelength resolution of 5 pm to obtain the transmission spectrum, as shown in Fig. 2(c). The free-spectral-range (FSR) was measured to be approximately 15.8 nm. Next, the Q factor of a mode around 1582.6 nm was measured by employing a fine scan at a wavelength resolution of 2 pm. The Q-factor of the mode was estimated to be $1.83 \times 10^5$ as indicated by the Lorentz fitting (red solid line) in Fig. 2(d), which is consistent with our previous results [2, 3, 6].

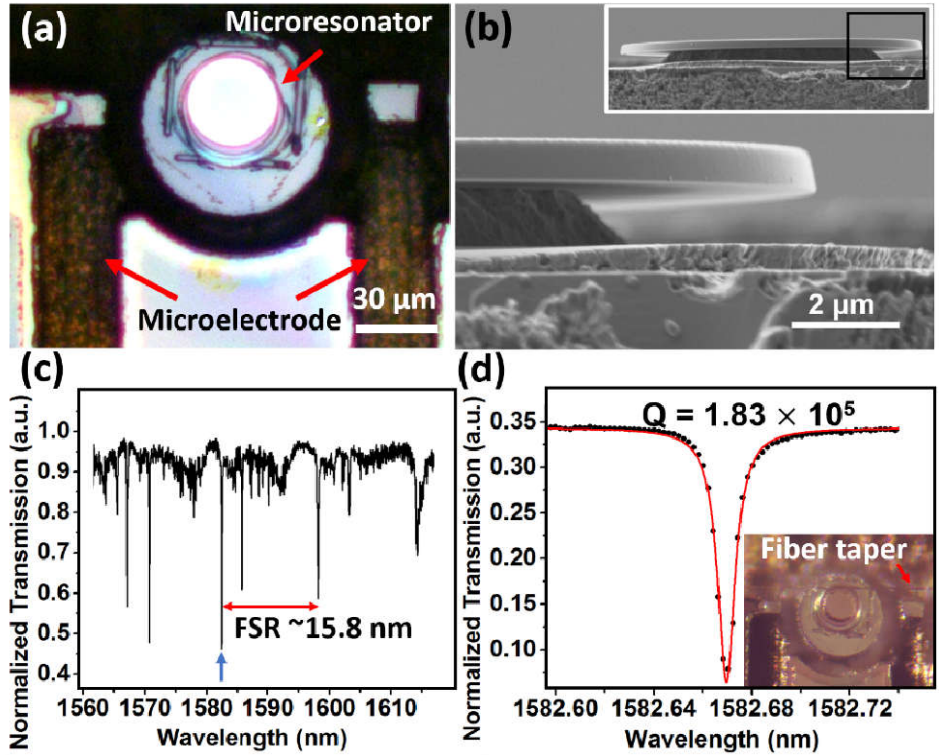

Fig. 2 (a) Top view of the EO tunable LN microresonator with integrated in-plane microelectrodes. (b) The SEM image of the freestanding LN microdisk with smooth boundary. Inset of (b) is the full image of the LN microdisk. (c) Normalized transmission spectrum of the LN microresonator measured under the critical coupling condition without applying any voltage. (d) The Lorentz fitting (red curve) of the dip under over-coupling condition showing a Q-factor of $1.83 \times 10^5$. The inset of (d) is the top-view optical image of the LN microresonator coupled with a fiber taper.

At last, we examined the performance of the fully integrated on-chip EO tunable LN microresonator by applying voltage to the two microelectrodes. The voltage was tuned in a range from 0 V to 150 V. Figure 3(a) shows eight transmission spectra of that particular mode near 1582.6 nm. The resonant wavelengths were measured as 1582.62157 nm, 1582.63678 nm, 1582.65715 nm, 1582.66964 nm, 1582.71716 nm, 1582.77198 nm, 1582.8258 nm, and 1582.89872 nm, with corresponding applied voltages of 0 V, 33.2 V, 56.8 V, 65.7 V, 100.6 V, 112.0 V, 135.0 V and 150.5 V, respectively. The experimental results indicate a maximum 277.15 pm wavelength shift when the voltage was increased from 0 V to 150.5 V. Figure 3(b) plots the wavelength shift as a function of the applied voltage. It reveals that resonant wavelength of the WGM possesses a nonlinear dependence on the voltage between the microelectrodes (i.e., the electric field strength in the microdisk). The microresonator shows a relatively low linear tuning coefficient of 0.826 pm/V at low voltages (i.e., < 80 V), which can be attributed to Pockels effect of linear EO dependence. However, when the applied voltage is above 80 V, the resonant wavelength shifts drastically to the longer wavelength with the increasing voltage, where the second order Kerr effect dominants due to the small inter-electrode separation [15]. The tuning coefficient is estimated to be 3.41 pm/V by linear fitting (the green dash dot line in Fig. 3(b)), which is significantly higher than the tuning coefficient of 0.826 pm/V achieved at the low voltages. The result shows that the integrated in-plane microelectrodes can provide efficient EO tuning due to the strong electric field generated between two microelectrodes with small gap.

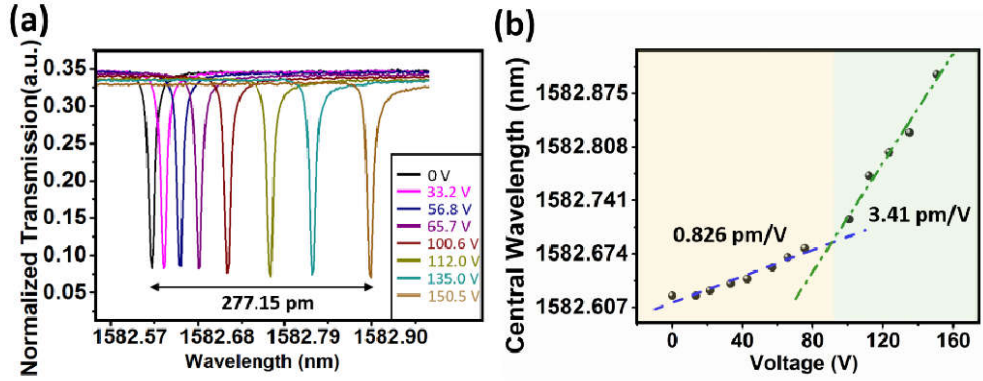

Fig. 3 (a) The spectrum near the resonant wavelength in the fabricated microresonator as a function of the applied voltage. (b) The resonant wavelength shift plotted as a function of the voltage indicates a nonlinear relationship (red solid line). At low voltages (i.e., 0 ~ 80 V), the tuning efficiency is 0.826 pm/V according to the linear fitting plotted by the blue dash line. At high voltages (i.e., above 80 V), the tuning coefficient is 3.41 pm/V according to the linear fitting plotted by the green dash dot line.

4. **Conclusion**

To conclude, we have demonstrated an on-chip tunable LN microresonator with in-plane integrated microelectrodes. Our technique relies on consecutive patterning of the microelectrodes and microresonators with high-precision femtosecond laser direct writing, which leads to high spatial precision sufficient for accurately assembling the isolated components into the fully integrated microdevice. It is also noteworthy that our electrodes are bulk metallic structures embedded within the LN substrate [13], which can produce strong bonding between the microelectrodes and the substrates. Furthermore, the electric field is more uniformly distributed in the plane perpendicular to the LN microdisk in comparison with conventional microelectrodes formed by patterning a metal thin film coated on the surface of LN substrate. Although we demonstrate electro-optic tuning of a single microresonator, our technique can be readily applied to fabricate coupled multiple microresonators whose resonant wavelengths can be tuned independently. Our technique shows promising potential for applications ranging from classical optical filters and modulators to cavity quantum electrodynamics.